# Skilful global seasonal predictions from a machine learning weather model trained on reanalysis data

Chris Kent[1*], Adam A. Scaife[1,2], Nick J. Dunstone[1], Doug Smith[1], Steven C. Hardiman[1], Tom Dunstan[1], Oliver Watt-Meyer[3]

1: Met Office Hadley Centre, Exeter, UK

2: Department of Mathematics and Statistics, University of Exeter, UK

3: Allen Institute for Artificial Intelligence (Ai2), Seattle, WA, USA

* corresponding author: chris.kent@metoffice.gov.uk

# Abstract

Machine learning weather models trained on observed atmospheric conditions can outperform conventional physics-based models at short- to medium-range (1-14 day) forecast timescales. Here we take the machine learning weather model ACE2, trained to predict 6-hourly steps in atmospheric evolution and which can remain stable over long forecast periods, and assess it from a seasonal forecasting perspective. Applying persisted sea surface temperature (SST) and sea-ice anomalies centred on 1st November each year, we initialise a lagged ensemble of winter predictions covering 1993/1994 to 2015/2016. Over this 23-year period there is remarkable similarity in the patterns of predictability with a leading physics-based model. The ACE2 model exhibits skilful predictions of the North Atlantic Oscillation (NAO) with a correlation score of 0.47 (p=0.02), as well as a realistic global distribution of skill and ensemble spread. Surprisingly, ACE2 is found to exhibit a signal-to-noise error as seen in physics-based models, in which it is better at predicting the real world than itself. Examining predictions of winter 2009/2010 indicates potential limitations of ACE2 in capturing extreme seasonal conditions that extend outside the training data. Nevertheless, this study reveals that machine learning weather models can produce skilful global seasonal predictions and heralds a new era of increased understanding, development and generation of near-term climate predictions.

# Introduction

In recent years a revolution in weather prediction has occurred in which machine learning-based models can match or outperform physics-based models over a range of metrics (Lam et al., 2023, Bi et al., 2023, Kurth et al., 2023, Chen et al., 2024, Price et al., 2025). Learning the 1–6-hour evolution of the atmospheric state, these models can produce skilful forecasts for several days by feeding the predictions back into themselves, known as "autoregressive" forecasting (de Burgh-Day & Leeuwenburg, 2023). Recent studies suggest skilful forecasts can be made covering several weeks (Ling et al., 2024, Guo et al., 2024) and very large ensembles can provide improved estimates of extreme events (Mahesh et al., 2024). Beyond these timescales, instabilities can grow, or the predictions blur and smooth, restricting their application to long-range climate predictions at monthly or seasonal time scales (Watt-Meyer et al., 2023, Karlbauer et al., 2024, Kochkov et al., 2024, Price et al., 2025, Yang et al., 2025).

Machine learning predictions at seasonal timescales often utilise more direct approaches in learning relationships between predictors and specific predictands, or resort to using model data for training. For example, skilful predictions have been demonstrated for the El Niño-Southern Oscillation (ENSO) and regional-scale climate variability (Ham et al., 2019, Qian et al., 2020, Kim et al., 2022, Taylor & Feng, 2022, Mu et al., 2023, Qian & Jia, 2023, Sun et al., 2024). Understanding the mechanisms underpinning such predictions can be difficult and developing methods to provide explainability is a key topic of research (Labe & Barnes, 2021, Eyring et al., 2024). With only one event per season, a key limitation at longer forecast periods is the relatively small sample size available for training. This restricts the ability to learn complex relationships whilst at the same time keeping a suitable number of years separate for testing, as needed for dynamical models (Manzanas et al., 2022). Alternatively, long-range predictions can use model data for training (Toms et al., 2021, Taylor & Feng, 2022, Pan et al., 2022) but the errors and biases found in physics-based models are inevitably inherited.

In this study we assess the newly developed machine learning weather model ACE2 (Watt-Meyer et al., 2024) from a seasonal forecasting perspective. This model predicts the atmospheric evolution at 6-hourly time steps and can remain stable for long autoregressive forecast periods, enabling it to provide seasonal simulations even though it was not explicitly trained to provide such predictions. It is trained only on historical conditions from the ERA5 dataset (Hersbach et al., 2020). We initialise ACE2 during autumn each year from 1993 to 2015 (see Methods) and assess the interannual skill of December-January-February (DJF) conditions. To provide boundary conditions, the SST and sea-ice anomalies at the time of initialisation are persisted throughout the forecast period each year. The interannual signals from large-scale drivers such as ENSO are therefore preserved, but any coupled ocean-atmosphere processes are missing. We compare the ACE2 seasonal forecasts to those from GloSea, a leading physics-based ensemble prediction system (Maclachlan et al., 2015, Kettleborough et al., 2025).

# Results

## Skilful data-driven seasonal forecasts

Over the 23-year assessment period it is striking how the pattern of interannual skill demonstrated by ACE2 closely resembles that of the dynamical model for mean sea level pressure (MSLP, Figure 1a). Whilst much of the tropical skill is due to the persistence of slowly evolving processes such as ENSO from the initialisation of the tropical oceans (Ehsan et al., 2024, Scaife et al., 2019), ACE2 also exhibits skill across the tropical land and the extratropics, including the North Atlantic and North Pacific (Figure 1ab). Interestingly, ACE2 also exhibits reduced skill over Eurasia, as seen in the physics-based model GloSea. In most regions the ACE2 correlation is weaker than that for GloSea. For example, the area-average correlation across the northern hemisphere extratropics (20°N to 90°N) is 0.39 in ACE2 and 0.44 in GloSea, whilst over the tropics (20°S to 20°N) the scores are 0.79 and 0.82 respectively.

For temperature (Figure 1c, d) we continue to see large regions of interannual skill from ACE2, including South America, Africa, Australia and parts of North America. As seen for MSLP, GloSea outperforms ACE2 across many parts of the world with the area-weighted mean correlation across the northern hemisphere extratropics at 0.41 in ACE2 and 0.45 in GloSea, and 0.68 and 0.77 respectively across the tropics. The interannual skill for both systems is lower for precipitation, however the ACE2 model (Figure 1e) once again closely resembles that of GloSea (Figure 1f), particularly across the tropics, the Caribbean and east Asia.

These results demonstrate that the ACE2 model can skilfully predict interannual variability across many parts of the world when initialised a month in advance.

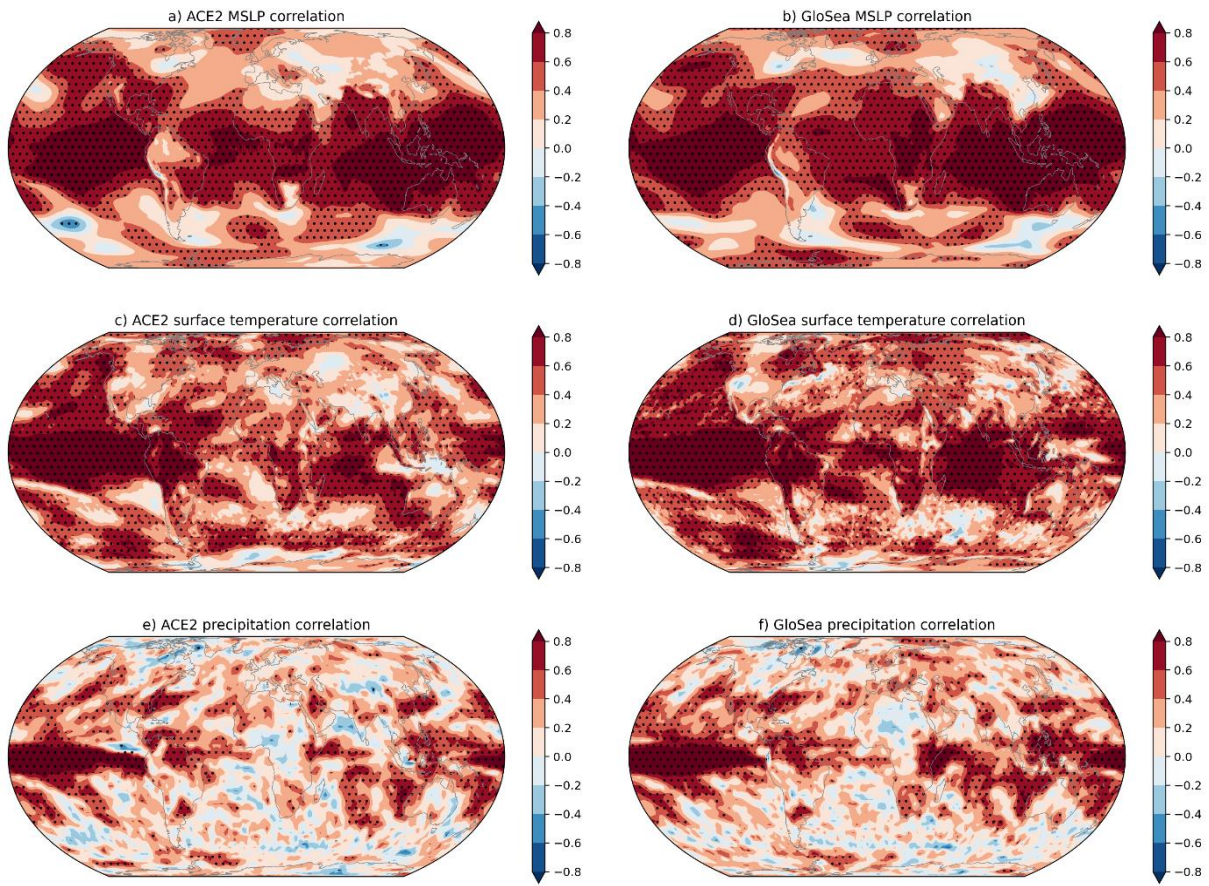

*Figure 1: Skilful interannual winter (DJF) predictions from the ACE2 machine learning and GloSea dynamical models. Correlation score of mean sea level pressure (a), near-surface temperature (c) and precipitation (e) for ACE2 and GloSea (b, d, f) calculated across winters 1993/1994 to 2015/2016. Stippling indicates significant correlations (23 years, 95% confidence level).*

## Predictability of the North Atlantic Oscillation

The NAO is the primary mode of interannual variability across the North Atlantic (Hurrell et al., 2003) and is a key focus for extratropical seasonal prediction (Scaife et al., 2014, Smith et al., 2016, Baker et al., 2024). ACE2 can predict the winter NAO (Stephenson et al., 2006, see Methods) with a correlation score of r=0.47 (Figure 2a). This skill is statistically significant at the 95% level (p=0.023) and is highly competitive with a range of dynamical models. For example, over a shorter 19-year analysis period (1993-2011) ACE2 exhibits higher NAO skill (r=0.42) than 4 operational ensemble prediction systems (Baker et al., 2024).

It is important to note that only the 9 winters between 2002-2010 are fully independent of the ACE2 training period (Watt-Meyer et al., 2024). Over this period the NAO correlation remains high (r=0.6), although with reduced significance due to the smaller sample size (p=0.07). Importantly, ACE2 gives a poor prediction of the extreme winter in 2009/2010 (see below). Nevertheless, given the long autoregressive forecasts, the lack of a well resolved stratosphere (Watt-Meyer et al., 2024), and the use of non-interacting, persisted SSTs, it is remarkable that the ACE2 model is able to exhibit such prediction skill for the NAO.

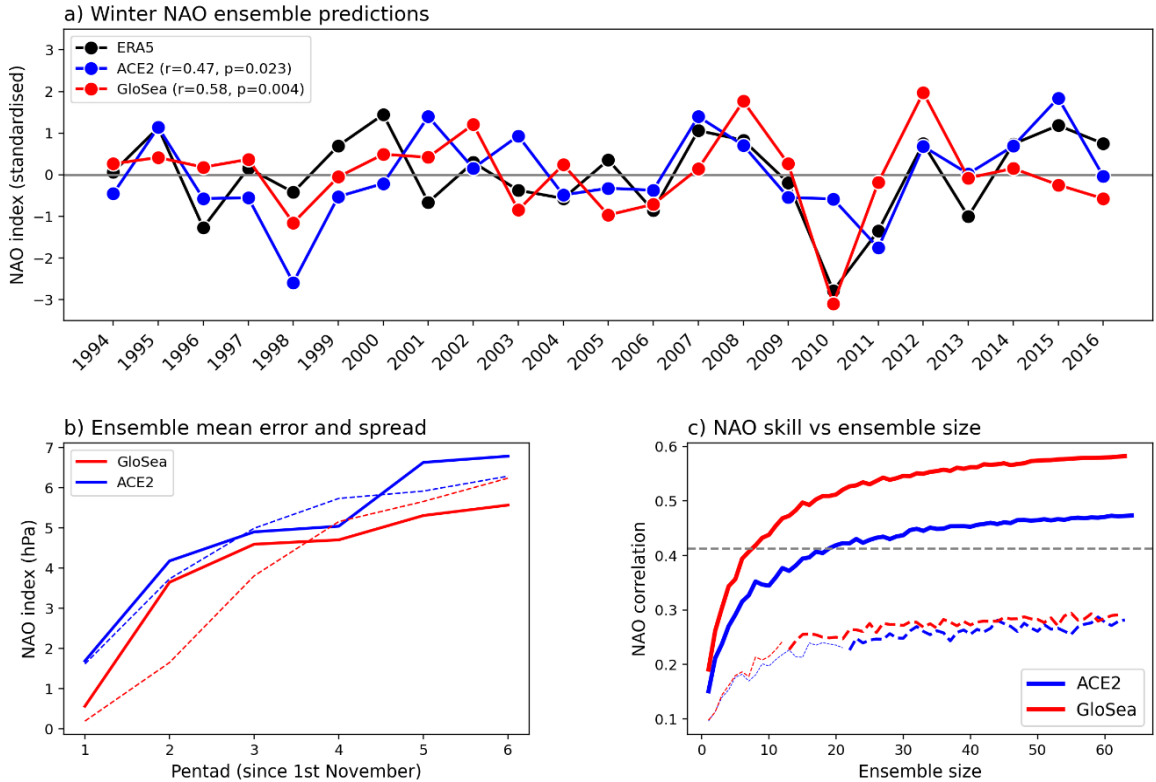

*Figure 2: Skilful predictions of the wintertime North Atlantic Oscillation (NAO). a) DJF mean NAO index, standardised to unit variance, from ERA5 (black), GloSea (red) and ACE2 (blue). b) ensemble mean RMSE (solid) and spread (dashed) for the NAO (hPa) as a function of lead time during November each year, averaged over all winters (see Methods). c) Relationship between NAO interannual correlation score and ensemble size (solid lines) and skill in predicting individual withheld ensemble members (dashed lines) based on 1000 random samples with no replacement. The dashed lines are thickened when significantly below the corresponding solid line (outside 95% sampling range). The horizontal dashed grey lines indicates the 95% significance level for a sample size of 23 years.*

We find that the ACE2 and GloSea NAO predictions are not strongly correlated (r=0.34, p=0.11) and so there may be additional value in combining them. Indeed, an ensemble mean constructed from both models results in an NAO correlation score of r=0.65 (p < 0.01), matching that estimated by GloSea with an extended ensemble size of 127 members. This indicates that ACE2 could also be utilised to enhance dynamical model ensembles.

In addition to skilful interannual predictions, the ACE2 ensemble closely matches the dynamical model in terms of NAO variability. Following initialisation, we find that the ACE2 ensemble mean error and ensemble spread increase in line with GloSea (Figure 2b). Furthermore, the winter mean total standard deviation across all years and members is 4.3 hPa in ERA5, 3.6 hPa in ACE2 and 3.8 hPa in GloSea. For the ensemble mean interannual variability the standard deviation is 1.11 hPa in ACE2 and 1.21 hPa in GloSea. The lagged-ensemble methodology used here therefore enables sufficient ensemble member spread to develop, but other methods for ensemble generation are key topics for future research.

In line with dynamical models (Scaife et al 2014, Dunstone et al. 2016, Baker et al., 2018), ACE2 NAO skill increases strongly with ensemble size (solid line, Figure 2c). This is encouraging as it is much cheaper and quicker, in computational terms, to increase the ensemble size of data-driven models compared to dynamical models. However, it can also be seen that when the ACE2 ensemble mean is used to predict one of its own individual members (so-called 'perfect model' skill), the skill is markedly lower (r ~ 0.25, dashed lines in Figure 2c) than the ACE2 skill in predicting the observed NAO (thick dashed lines, Figure 2c). The ratio of predictable components for ACE2 (see Methods) is found to be 1.6, only slightly less than the 1.8 for GloSea, but still greater than 1 (90% confidence). Somewhat surprisingly therefore, despite having been trained only on reanalysis data, the ACE2 predictions also exhibit a signal-to-noise error which resembles that found in dynamical models (Eade et al., 2014, Scaife et al., 2014, Dunstone et al., 2016, Scaife & Smith 2018, Weisheimer et al., 2024). This may suggest that the signal-to-noise error is not due to a physical model error and instead occurs due to some other damping effect on the predictable signal. Machine learning predictions can exhibit such damping, particularly in terms of the kinetic energy spectrum (Bonavita, 2024, Karlbauer et al., 2024), which in principle could give rise to characteristics similar to the signal-to-noise error. Therefore, it is possible that the same qualitative behaviour occurs for different reasons in the ACE2 and GloSea models, but further research is needed to understand if this is the case.

## ENSO as a driver of interannual skill

ENSO is the primary mode of interannual climate variability and is a key driver of seasonal skill across many parts of the world (Horel & Wallace, 1981, Taschetto et al., 2020). In this section we investigate whether ACE2 is correctly capturing ENSO teleconnections.

Figure 3Composite differences between El Niño and La Niña years (Figure 3) reveal that ACE2 exhibits very similar teleconnection patterns to those seen in ERA5 and GloSea for both MSLP and surface temperature. In particular, we find El Niño deepens the Aleutian low and influences the North Atlantic jet, extending eastward from the Caribbean. This suggests that ACE2 is capturing the ENSO relationship on the subtropical jet, an important mechanism underpinning the global influence of ENSO (e.g. Horel & Wallace, 1981, Jiménez-Esteve & Domeisen, 2018). In terms of the surface temperature response, ACE2 also exhibits very similar ENSO teleconnections to ERA5 and GloSea, particularly over North America, South America, southern Africa and Australia. These composites indicate that ACE2 is correctly capturing the regional interannual variability associated with ENSO across many parts of the world despite being trained only on the 6-hourly evolution of the atmosphere (Watt-Meyer et al., 2024).

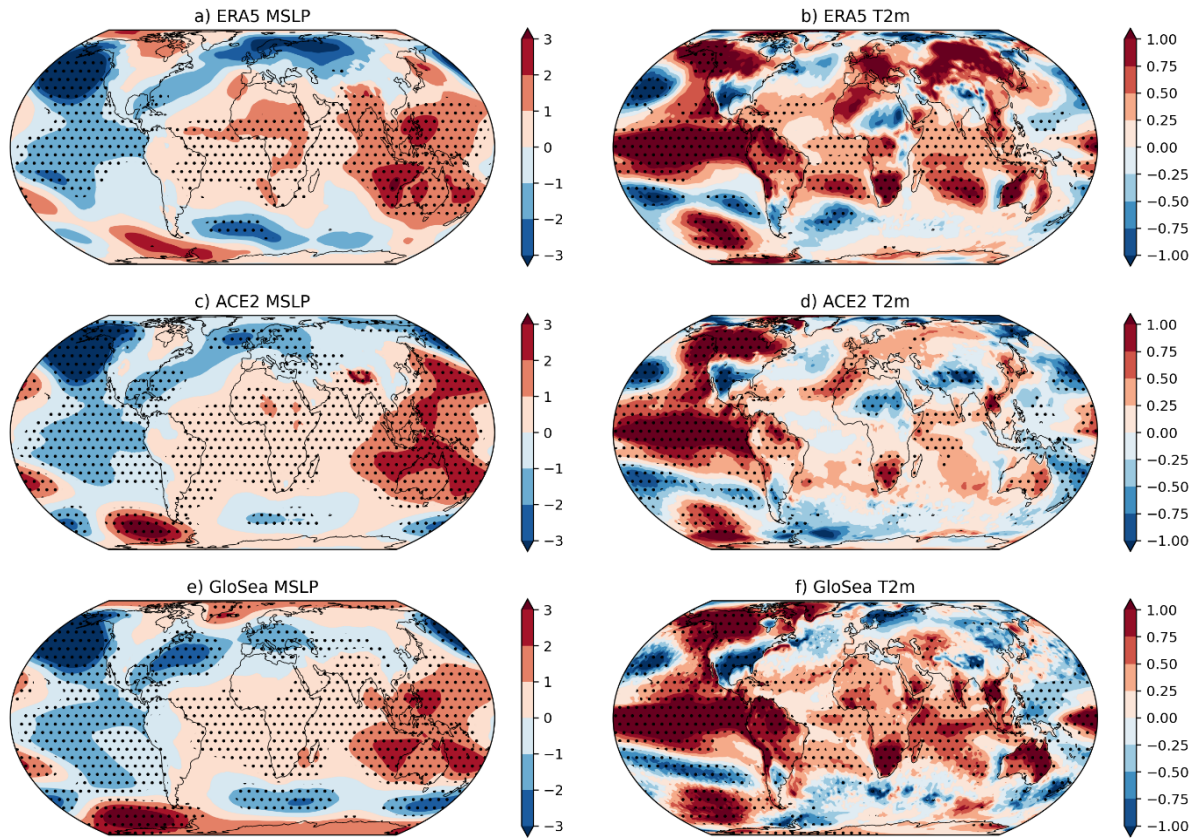

*Figure 3: Influence of ENSO on winter surface conditions. Composite maps of El Niño years (n=8) minus La Niña years (n=9) for mean sea level pressure (hPa) and surface temperature (K) anomalies for ERA5 (ab), ACE2 (cd) and GloSea (ef). Shaded contours show the DJF mean anomaly. Stippling indicates significant differences (two-tailed T-test, 95% confidence level).*

## The extreme winter of 2009/2010

As a final part of our assessment we focus on predictions for the extreme winter of 2009/2010, which is part of the independent dataset withheld during the training of ACE2. This winter is characterised by a record negative NAO, well beyond the anomalies seen in other years. It was also subject to a minor and a major sudden stratospheric warming (SSW), a strong El Niño and an easterly Quasi Biennial Oscillation (QBO, Fereday et al., 2012).

The winter mean MSLP anomaly (Figure 4a) exhibits a very zonal negative NAO which is well captured by GloSea (Figure 4c). However, the ACE2 ensemble mean prediction does not appear to capture this signal with only slightly above average pressure across the Arctic (Figure 4b). This is surprising given the strong tropical forcing and potentially indicates a limitation of ACE2 in predicting extreme, out of sample conditions. Exploring this further, we find that both ERA5 and GloSea exhibit a weakened stratospheric polar vortex (Figure 4de), whilst ACE2 exhibits near-normal vortex strength (Figure 4f).

In terms of SSWs, the winter comprised of a minor warming in December 2009 and a major warming in January 2010, reflecting the increased SSW probability due to the El Niño and easterly QBO (Fereday et al 2012, Garfinkel et al., 2012, Domeisen et al., 2019, Anstey et al., 2022). GloSea appears to capture this increase, with 81% of members (51 out of 63) experiencing easterly zonal winds at 10hPa and 60°N within the winter. This is significantly higher than GloSea's climatological probability of 62% (two proportion Z-test, 95% confidence level). In comparison, only 39% of ACE2 members (25 out of 64) exhibit easterly stratospheric winds in the upper most model layer (above 50mb), which is not significantly different to the climatological rate of 40%. This indicates that the ACE2 model is not correctly capturing the disruption to the stratospheric polar vortex during winter 2009/2010.

Furthermore, the SSW probability within ACE2 is relatively consistent across El Niño (45%) and La Niña (36%) years, neither of which are significantly different from neutral years (41%, one-tailed two proportion Z-test, 95% confidence level). GloSea and ERA5 however exhibit significant differences between active and neutral ENSO years, with a higher chance of an SSW during El Niño (Bell et al 2009, Butler & Polvani, 2011, Bett et al., 2023, Ineson et al., 2024). This suggests that whilst the ACE2 can exhibit sub-seasonal stratospheric variability it is not fully capturing the ENSO teleconnection to the stratosphere despite realistic tropospheric teleconnections (Watt-Meyer et al., 2024).

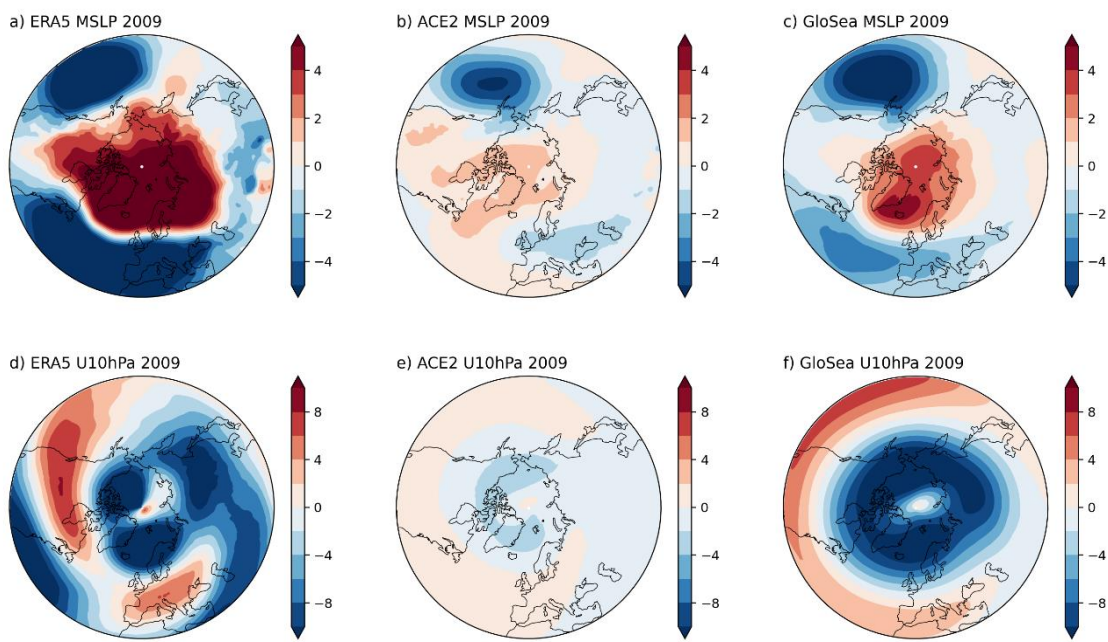

*Figure 4: Surface and stratospheric anomalies associated with the extreme winter of 2009/2010. Anomalies from the 1994-2016 climatology of MSLP (hPa) and zonal wind at 10hPa (ms$^{-1}$) for ERA5 (a, d), ACE2 (b, e) and GloSea (c, f). ACE2 stratospheric conditions are model layer 0 (above 50 hPa).*

# Discussion and conclusions

This study demonstrates skilful seasonal predictions from a machine learning weather model. Despite being trained only on the 6-hourly observed evolution of the atmosphere, when assessed from a seasonal prediction perspective, the ACE2 model exhibits significant interannual skill and is competitive with current dynamical systems. A lagged-ensemble approach is found to generate an ensemble spread which closely match observations and a physics-based ensemble prediction system, a characteristic is it not specifically trained on. The model produces realistic ENSO teleconnections in the troposphere, but the stratospheric pathway is not in line with observations. This may be due to a relatively small sample of observed events, the training methodology (e.g. the loss weightings applied to different levels or parameters), or model architecture. If the latter, this could potentially be addressed through enhanced vertical resolution in the stratosphere, a characteristic found to be important in dynamical models (Bell et al., 2009, Ineson & Scaife, 2009, Cagnazzo & Manzini 2009, Butler et al., 2016), providing an opportunity for improved skill in the future.

Whilst part of the assessed period falls within the training sample for ACE2, we find little evidence for reduced performance in the solely independent years. Furthermore, with such long autoregressive rollouts and persisted boundary conditions as utilised here, the model must generate its own trajectory throughout several months and yet still exhibits suitable ensemble spread. This provides confidence that the results are not dependent on the overlapping periods.

A significant benefit of data-driven models is the relatively cheap computational cost. For seasonal forecasting timescales, a dynamical model can take hours on a supercomputer per simulation. In comparison, the ACE2 model can complete a 4-month forecast simulation in under 2 minutes on an Nvidia A100 GPU. Opportunities arising from this include the ability to generate very large ensemble sizes (e.g. over 7000 members in Mahesh et al., 2024), much longer assessment periods, rapid testing of new experimental setups and better exploration of sources of predictability and the signal-to-noise error (Weisheimer et al., 2024). Data-driven models are therefore highly applicable for seasonal and climate timescales where large ensembles are needed. In particular, further research is needed on optimal ensemble generation approaches as well as coupling to data-driven ocean models (e.g. Clark et al., 2024) or ocean-atmosphere-coupled dynamical models. However, it is clear from this work that the machine learning models can supplement and support current seasonal forecasting methods.

Overall, these results show that the machine learning revolution is not limited to short-range weather forecasts and raises the prospect of a new era for near-term climate predictions.

# Methods

## Datasets

Historical atmospheric conditions for the period 1993-2016 are taken from the ERA5 reanalysis (Hersbach et al., 2020). To persist SST and sea-ice conditions throughout a forecast we create a seasonally varying climatology based on the 6-hourly atmospheric state, for each grid cell, using a rolling-mean gaussian filter with a width (standard deviation) of 10 days. Observed monthly rainfall totals are taken from the Global Precipitation Climatology Project version 2.3 (GPCP, Adler et al. 2003).

For comparison with dynamical models, hindcasts (retrospective forecasts) covering winters 1994 – 2016 are taken from the GloSea operational ensemble prediction system with GC3.2 configuration (MacLachlan et al., 2015, Williams et al., 2018, Kettleborough et al., 2025). A 63-member ensemble is constructed from 21 members initialised on 25$^{th}$ October, 1$^{st}$ November and 9$^{th}$ November each year.

For this study we use the machine learning atmospheric model ACE2 (Watt-Meyer et al., 2024). The model is trained solely on ERA5 reanalysis atmospheric fields and predicts the evolution of the atmospheric state at 6-hour time steps at a 1° grid resolution. Importantly, ACE2 autoregressive forecasts are stable over multiple years due to its Spherical Fourier Neural Operators architecture (SFNO, Bonev et al., 2023), use of user prescribed ocean and sea-ice boundary conditions, and physical constraints on mass conservation, moisture, precipitation rate and radiative fluxes (Watt-Meyer et al., 2024).

Full details of the ACE2 training are provided in Watt-Meyer et al. (2024). Of relevance to this study, the 10 years from 2001-2010, which lies within our 23-year hindcast period, are withheld and form an independent test period for the model. The remaining winters are used to train the model. However, our experiments (see below) are initialised one month prior to the periods of interest and utilise persisted boundary conditions. These specific atmospheric and ocean states will therefore be new to the model, although the large-scale patterns will have been seen previously. In addition, the model must calculate its own trajectory from the initialised state to the end of forecast period (i.e. long autoregressive rollouts) and thus can still be evaluated for skill with some confidence over the entire hindcast period.

All ERA5 and GloSea data is bilinearly interpolated to the native 1° x 1° ACE2 grid, except for precipitation, in which ACE2 and GloSea are interpolated to the 2.5° x 2.5° GPCP grid.

## Indices and metrics

We define ENSO years based on the DJF Oceanic Niño Index (ONI, NOAA/CPC, 2025) and a threshold of ±0.5 K. El Niño winters are 1995, 1998, 2003, 2005, 2007, 2010, 2015 and 2016. La Niña winters are 1996, 1999, 2000, 2001, 2006, 2008, 2009, 2011 and 2012.

We define the NAO index following Stephenson et al. (2006) as the difference in mean sea level pressure between a southern (90°W–60°E, 20°N–55°N) and a northern box (90°W–60°E, 55°N–90°N). The results are consistent when applying a smaller regional definition (r=0.42, p=0.048, Dunstone et al., 2016) and a point-based estimate (r=0.41, p=0.053, Scaife et al., 2014).

To assess ACE2 and GloSea predictions in terms of signal and noise, we compute the ratio of predictable components (RPC, Eade et al., 2014) as:

$$RPC = \frac{r}{\sigma_{sig}/\sigma_{tot}}$$

Where r is the ensemble mean correlation with ERA5, $\sigma_{sig}$ is the ensemble mean standard deviation and $\sigma_{tot}$ is the standard deviation across all members and years. A random resampling procedure (Eade et al., 2014) is used for significance testing.

To calculate the ensemble mean error and spread as a function of lead time (Figure 2b) we utilise only ACE2 members initialised between 00:00z on 27$^{th}$ October and 18:00z on 31$^{st}$ October (n=20) each year and GloSea members initialised at 00:00z on 1$^{st}$ November (n=21). Daily NAO index values are aggregated into pentads and the climatological mean removed. The ACE2 values are therefore partly larger than GloSea's due to the inclusion of longer lead time forecasts. The ensemble mean error for a given pentad (5-day average), RMSE$_p$ is defined as:

$$RMSE_p = \sqrt{\frac{1}{23} \sum_{i=1994}^{2016} (model_{i,p} - ERA5_{i,p})^2}$$

The corresponding average ensemble spread σ$_p$ is defined as:

$$\sigma_p = \sqrt{\frac{1}{23} \sum_{i=1994}^{2016} \sigma_{ip}^2}$$

Where $\sigma_{ip}$ is the standard deviation of the model NAO across members for year i and pentad p.

## ACE2 experimental setup

ACE2 seasonal predictions are generated using a lagged ensemble approach. An ensemble member is initialised every 6 hours between 25$^{th}$ October and 9$^{th}$ November each year from 1993 to 2015, creating a total of 64 members per winter. The forecast period extends from initialisation through to mid-March the following year. Initial conditions for each member are taken from the ERA5 reanalysis dataset (Hersbach et al., 2020).

Boundary SST and sea-ice conditions are provided throughout each forecast by calculating the instantaneous anomaly at initialisation for each grid cell and persisting this throughout the winter using the derived ERA5 6-hourly climatology. Historical downward shortwave radiative flux at top of the atmosphere and global mean atmospheric carbon dioxide inputs prescribed throughout the hindcast period and taken from Watt-Meyer et al. (2024). Repeating the hindcast experiment whilst either increasing (or decreasing) the annual $CO_2$ value by a constant 2 ppm produces relatively consistent results although the NAO correlation score is reduced. However, these are in line with results obtained from a separate hindcast assessment in which the original initial condition times are manually altered by 6 hours (i.e. to assess internal variability). GloSea hindcast simulations utilise annually repeating climatologies.

# Acknowledgements

This work was supported by the Met Office Hadley Centre Climate Programme (HCCP), funded by the UK Department for Science, Innovation and Technology (DSIT). The authors also thank Rowan Sutton, David Walters and Christopher Bretherton for their comments and feedback.

# Data availability

Initial conditions for ensemble members are taken from the ERA5 reanalysis dataset (Hersbach et al., 2020, see methods). $CO_2$ and solar irradiance forcing data are available at https://huggingface.co/allenai/ACE-2-ERA5. The data used for the figures is available at 10.5281/zenodo.15025230.

## Code availability

The trained ACE2-ERA5 model checkpoint used in this study is available at https://huggingface.co/allenai/ACE-2-ERA5.

## Contributions

C.K. performed the ACE2 seasonal predictions and carried out the analysis against GloSea. C.K., A.A.S, N.D., D.S. and S.H designed the experimental set up and interpreted the results. All authors contributed to the writing of the manuscript.

and 3.1 (GC3. 0 and GC3. 1) configurations. *Journal of Advances in Modeling Earth Systems*, *10*(2), pp.357-380.

Yang, S., Nai, C., Liu, X., Li, W., Chao, J., Wang, J., Wang, L., Li, X., Chen, X., Lu, B. and Xiao, Z., 2025. Generative assimilation and prediction for weather and climate. *arXiv preprint arXiv:2503.03038*.